\documentclass[a0,portrait]{a0poster}

\usepackage[absolute]{textpos}

\usepackage{graphics,wrapfig,times}

\usepackage{color}
\textblockcolour{white}

\def\LHead#1{\noindent{\Large #1}\smallskip}
\def\Title#1{\noindent{\VeryHuge #1}}

\TPGrid[40mm,40mm]{15}{25}  

\usepackage{wallpaper}

\begin{document}

\begin{textblock}{15}(0,0)
  \begin{center}
    \baselineskip=3\baselineskip \Title{Import of ENZYME data into the ConceptWiki\\and its representation as RDF}
  \end{center}
\end{textblock}

\begin{textblock}{15}(0,1.5)
  \begin{center}
    \LHead{Paul Boekschoten, Kees Burger, Barend Mons, Christine Chichester}
  \end{center}
\end{textblock}

\begin{textblock}{7}(0.1,2.5)
  \resizebox{0.3\TPHorizModule}{!}{\includegraphics{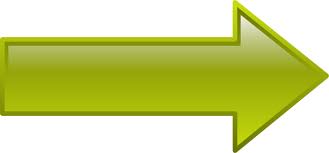}}
  \LHead{Abstract}

\indent{Solutions to the classic problems of dealing with heterogeneous data and making entire collections interoperable while ensuring that any annotation, which includes the recognition-and-reward system of scientific publishing, need to fit into a seamless beginning to end to attract large numbers of end users. The latest trend in Web applications encourages highly interactive Web sites with rich user interfaces featuring content integrated from various sources around the Web. The obvious potential of RDF, SPARQL, and OWL to provide flexible data modeling, easier data integration, and networked data access may be the answer to the classic problems. Using Semantic Web technologies we have created a Web application, the ConceptWiki, as an end-to-end solution for creating browser-­based read­‐write triples using RDF, which focus on data integration and ease of use for the end user. Here we will demonstrate the integration of a biological data source, the ENZYME database, into the ConceptWiki and it's representation in RDF.}
\end{textblock}

\begin{textblock}{7}(0.1,12.5)

The ConceptWiki (www.conceptwiki.org, Figure 1)	is an open access repository of editable concepts. It is a web based wiki system that accepts essentially unlimited numbers of synonyms, in multiple languages, and then maps all the terms correctly back to one unique concept identifier, alleviating problems of vocabulary and identifier differences. Each concept in the ConceptWiki is annotated with one or more semantic types and basic information like a definition. Users can view and edit information through a uniform interface. The information in the system is stored and edited in a highly structured way, as triples (e.g. <concept A> <has synonym> <term B>). The ConceptWiki backend has been designed to support the storage of concepts in a very generic form, thereby trying to avoid as much as possible the exclusion of potential valuable information sources. This compatibility with our other information storage systems enables higher­‐level applications to easily query, summarize and mine the knowledge. In line with recommendations from the Concept Web Alliance, identifiers in the ConceptWiki are completely opaque; they have no inherent structure and no information can be derived from them. An opaque identifier is a robust identifier as there will never be a need to change the identifier when underlying information changes. The ConceptWiki uses Universally Unique Identifiers (UUIDs) for identifiers.
\end{textblock}

\begin{textblock}{7}(8,3)

The ConceptWiki contains the biomedical terminology of Unified Medical Language System thesauri, concepts from SwissProt, Medline, and in the near future the repository will be expanded to incorporate the chemical terminology from ChemSpider for biologically relevant chemical molecules. In this poster, we demonstrate the import process for another biologically relevant database, ENZYME (Figure 2). For this import, the ENZYME flat file is first converted into XML. The import script incorporates a XML parser which queries the ConceptWiki database to recover all concepts that match those in the XML. For the ENZYME data, these are concepts containing an EC number as a synonym. The stored ConceptWIki information is then compared to the ENZYME data. If the ENZYME data is unknown to the ConceptWiki, then it is inserted into the database. If the ENZYME data is found but has changes in comparison to the ConceptWiki, then the stored information is updated. These changes are shown in the interface by removing the authority checkbox if data are no longer supported by ENZYME, and if the data are new the checkbox for ENZYME authority is added. EC 1.1.1.1 has synonym Aldehyde reductase, is an example of the ENZYME data generated in the subject-predicate-object triple structure which in XML is the triple structure used by the ConceptWiki backend. Each element is stored with a UUID.	
\end{textblock}

\begin{textblock}{7}(8,6.8)

The RDF data model (Figure 3) is similar to classic conceptual modeling approaches as it is based upon the idea of making statements about resources, in particular Web resources, in the form of subject-predicate-object expressions. The predicates of RDF triples are similar to hyperlinks; however, the advantage of RDF triples over HTML hyperlinks is that the links are explicitly labeled. The semantics of the relationship between the two entities is computationally accessible through URI resolution and of particular interest; the data for the ConceptWiki predicate is represented in RDF using the concept UUID. Of course, as would be expected, the subject and object are also represented in RDF using the ConceptWiki UUIDs.
\end{textblock}

\begin{textblock}{7}(8,12.5)

Using the concepts present in the ConceptWiki as a set of basic building blocks, new triples can be assembled by end users via the interface. Through the simple drop down menus, users can establish links between two records to illustrate pertinent connections. All newly created triples are attributed to the participating scientist by showing their name and listing the triple on page that represents them. For example, the triple ‘Aldehyde reductase has function sorbitol biosynthetic process’ can be built using data included from the ENYME import (Figure 4). Certain agencies have indicated that they are interested in using the resulting contributions as indices of scholarly achievement.
\end{textblock}

\begin{textblock}{7}(8,22)
  \resizebox{0.3\TPHorizModule}{!}{\includegraphics{yellowarrow.jpeg}}
  \LHead{Conclusion}
  
The ConceptWiki is well-equipped to make the task of generating new and more powerful Web mash-ups simple, elegant, and enjoyable for end users. Acknowledgement of individual contributions in the ConceptWiki interface will make initial publishing efforts more accessible to groups that are currently under-represented groups in the life sciences. While, the RDF output for a concept is fully linked to the related concepts in other triples, making the ConceptWiki a semantic network of concepts suitable for machines as well.
\end{textblock}

\begin{textblock}{3}(0.1,6)
  \begin{center}
    \setlength\fboxsep{0pt}
    \setlength\fboxrule{0.5pt}
    \fbox{\resizebox{7\TPHorizModule}{!}{\includegraphics{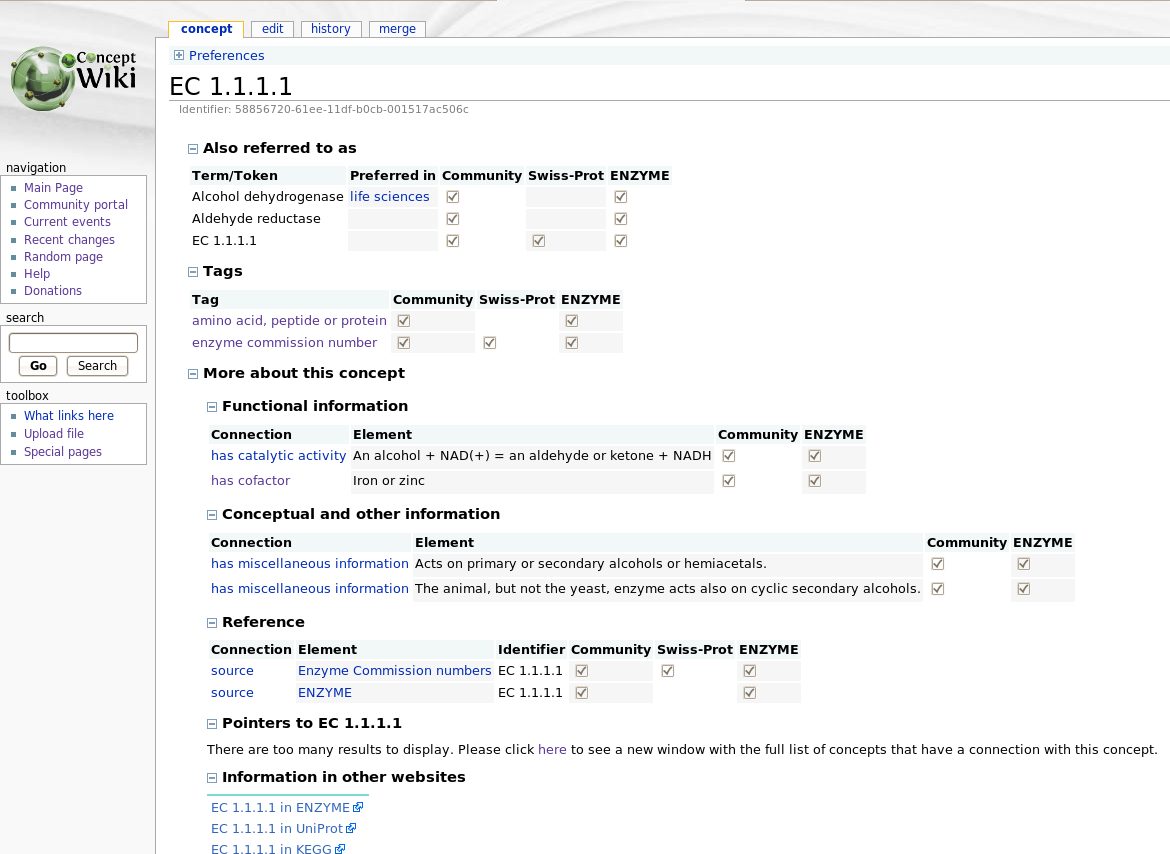}}}
    \\Figure 1: ConceptWiki
  \end{center}
\end{textblock}

\begin{textblock}{3}(0.1,16.5)
  \begin{center}
    \setlength\fboxsep{0pt}
    \setlength\fboxrule{0.5pt}
    \fbox{\resizebox{7\TPHorizModule}{!}{\includegraphics{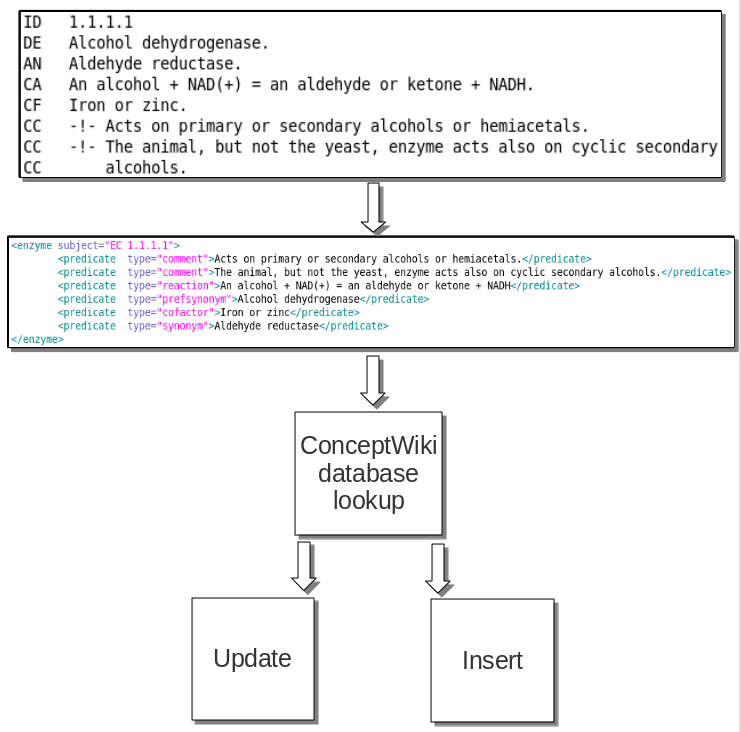}}}
    \\Figure 2: ENZYME data import
  \end{center}
\end{textblock}

\begin{textblock}{3}(8,9)
  \begin{center}
    \setlength\fboxsep{0pt}
    \setlength\fboxrule{0.5pt}
    \fbox{\resizebox{7\TPHorizModule}{!}{\includegraphics{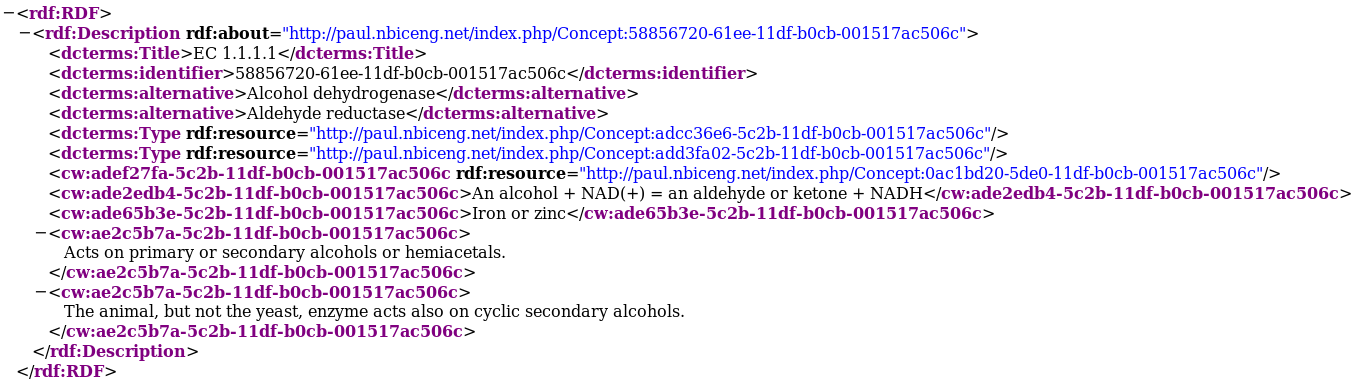}}}
    \\Figure 3: RDF of ENZYME data
  \end{center}
\end{textblock}

\begin{textblock}{3}(8,14.5)
  \begin{center}
    \setlength\fboxsep{0pt}
    \setlength\fboxrule{0.5pt}
    \fbox{\resizebox{7\TPHorizModule}{!}{\includegraphics{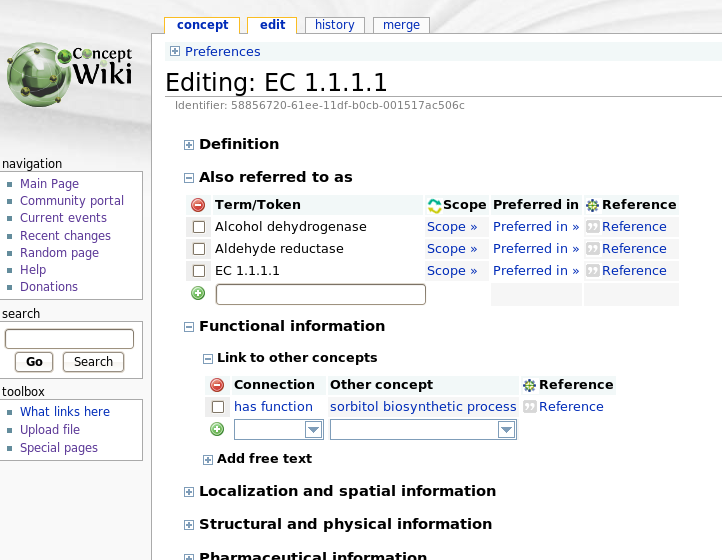}}}
    \\Figure 4: Building new triples with ENZYME data
  \end{center}
\end{textblock}

\LLCornerWallPaper{1}{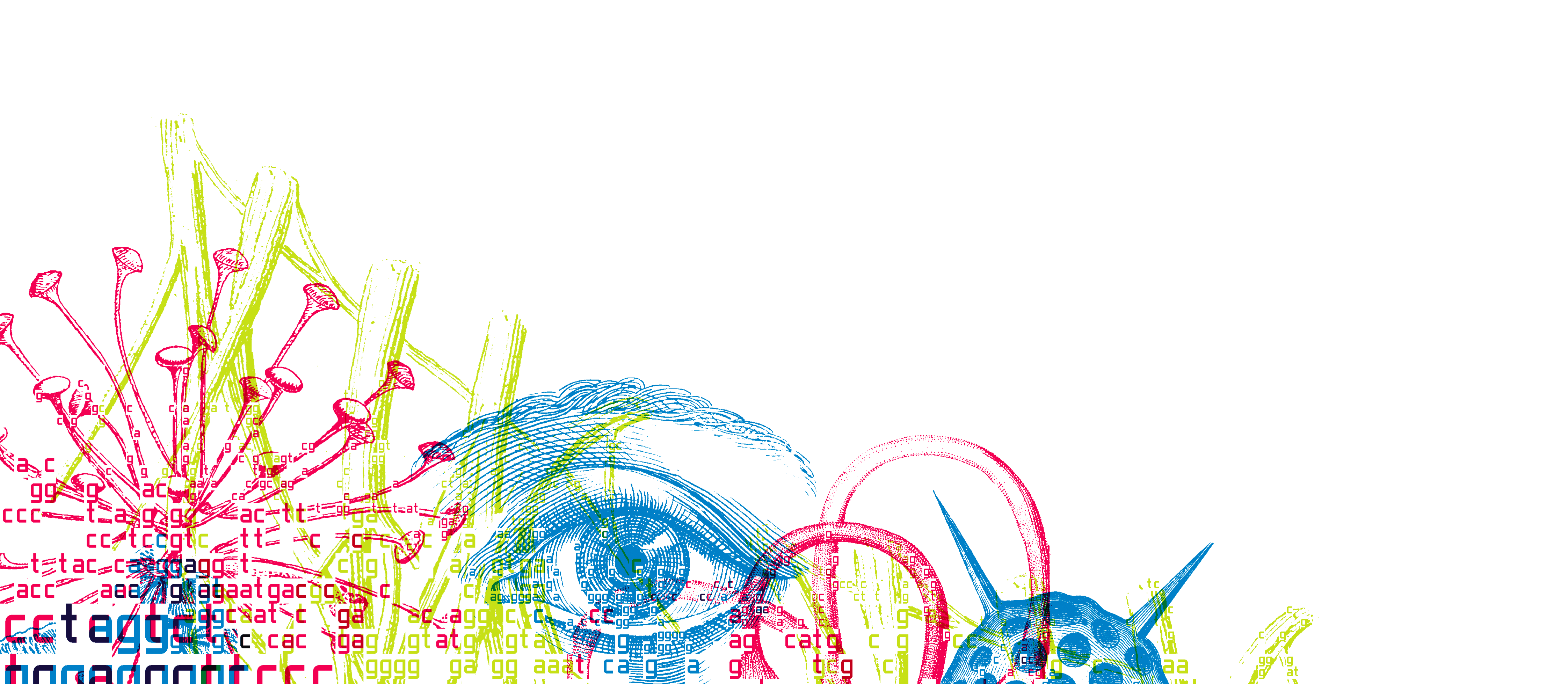}

\begin{textblock}{3}(12,24)
  \begin{center}
    \resizebox{3\TPHorizModule}{!}{\includegraphics{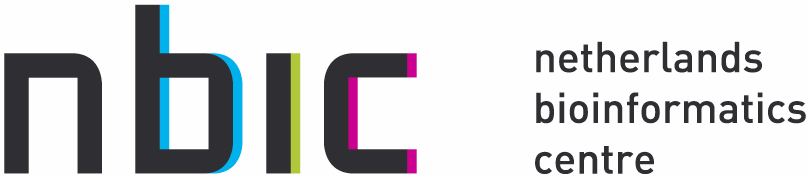}}
    \\http://www.nbic.nl
  \end{center}
\end{textblock}

\end{document}